# OPTICAL LINK OF THE ATLAS PIXEL DETECTOR

K.K. Gan, W. Fernando, P.D. Jackson, M. Johnson, H. Kagan,
A. Rahimi, R. Kass, S. Smith

*Department of Physics, The Ohio State University,*
*Columbus, OH 43210, USA*

P. Buchholz, M. Holder, A. Roggenbuck, P. Schade, M. Ziolkwoski

*Fachbereich Physik, Universitaet Siegen,*
*57068 Siegen, Germany*

The on-detector optical link of the ATLAS pixel detector contains radiation-hard receiver chips to decode bi-phase marked signals received on PIN arrays and data transmitter chips to drive VCSEL arrays. The components are mounted on hybrid boards (opto-boards). We present results from the irradiation studies with 24 GeV protons up to 32 Mrad ($1.2 \times 10^{15}$ p/cm$^2$) and the experience from the production.

## 1. Introduction

The ATLAS pixel detector [1] consists of three barrel layers, three forward, and three backward disks which provide at least three space point measurements. The pixel sensors are read out by front-end electronics controlled by the Module Control Chip (MCC). The low voltage differential signal (LVDS) from the MCC is converted by the VCSEL Driver Chip (VDC) into a single-ended signal appropriate to drive a VCSEL (Vertical Cavity Surface Emitting Laser). The optical signal from the VCSEL is transmitted to the Readout Device (ROD) via a fiber.

The 40 MHz beam-crossing clock from the ROD, bi-phase mark (BPM) encoded with the command signal to control the pixel detector, is transmitted via a fiber to a PIN diode. This BPM encoded signal is decoded using a Digital Opto-Receiver Integrated Circuit (DORIC). The clock and command signals recovered by the DORIC are in LVDS form for interfacing with the MCC.

The ATLAS pixel optical link system contains 632 VDCs and 544 DORICs with each chip containing four channels. The chips are mounted on 272 chip carrier boards (opto-boards). Each opto-board contains seven active optical links. The silicon components (VDC/DORIC and PIN) of the optical link will be exposed to a maximum total fluence of $3.7 \times 10^{14}$ 1-MeV n$_{eq}$/cm$^2$ during ten years of operation at the LHC. The corresponding fluence for the GaAs component (VCSEL) is 2 x $10^{15}$ 1-MeV $n_{eq}/cm^2$. We study the response of the optical link to a high dose of 24 GeV protons. The expected equivalent fluences at LHC are 6.3 and 3.8 x $10^{14}$ p/cm$^2$, respectively. In this paper we describe the results from the irradiations and the experience from production.

## 2. VDC Circuit

The VDC is used to convert an LVDS input signal into a single-ended signal appropriate to drive a VCSEL in a common cathode array. The output current of the VDC should be variable between 0 and 20 mA through an external control current, with a standing (dim) current of ~1 mA to improve the switching speed of the VCSEL. The nominal operating current for a VCSEL is 10 mA. The electrical output should have rise and fall times (20–80%) between 0.5 and 2.0 ns; nominally 1.0 ns. In order to minimize the power supply noise on the opto-board, the VDC should also have constant current consumption independent of whether the VCSEL is in the bright (on) or dim (off) state.

Figure 1 shows a block diagram of the VDC circuit. An LVDS receiver converts the differential input into a single-ended signal. The driver controls the current flow from the positive power supply into the anode of the VCSEL. The VDC circuit is therefore compatible with a common cathode VCSEL array. An externally controlled voltage ($V_{Iset}$) determines the current $I_{set}$ that sets the amplitude of the VCSEL current (bright minus dim current), while another externally controlled voltage (Tunepad) determines the dim current. The driver contains a dummy driver circuit which, in the VCSEL dim state, draws an identical amount of current from the positive power supply as is flowing through the VCSEL in the bright state. This enables the VDC to have constant current consumption.

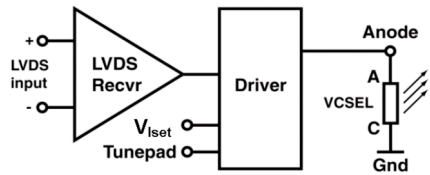

Figure 1. Block diagram of the VDC circuit.

## 3. DORIC Circuit

The DORIC decodes BPM encoded clock and data signals received by a PIN diode. The BPM signal is derived from the 40 MHz beam-crossing clock by sending transitions corresponding to clock leading edges only. In the absence of data bits (logic level 0), the resulting signal is a 20 MHz square wave. Data bits are encoded as an extra transition at beam-crossing clock trailing edges.

The amplitude of the current from the PIN diode is expected to be in the range of 40–1000 µA. The 40 MHz clock recovered by the DORIC is required to have a duty cycle of (50 ± 4)% with a total timing error (jitter) of less than 1 ns. The bit error rate of the DORIC is required to be less than $10^{-11}$ at end of life.

Figure 2 shows a block diagram of the DORIC circuit. In order to keep the PIN bias voltage (up to 15 V) off the DORIC, we employ a single-ended preamp circuit to amplify the current produced by the PIN diode. Since single-ended preamp circuits are sensitive to power supply noise, we utilize two identical preamp channels: a signal channel and a noise cancellation channel. The signal channel receives and amplifies the input signal from the anode of the PIN diode, plus any noise picked up by the circuit. The noise cancellation channel amplifies noise similar to that picked up by the signal channel. This noise is then subtracted from the signal channel in the differential gain stage. To optimise the noise subtraction, the input load of the noise cancellation channel should be matched to the input load of the signal channel (PIN capacitance) via an external dummy capacitance.

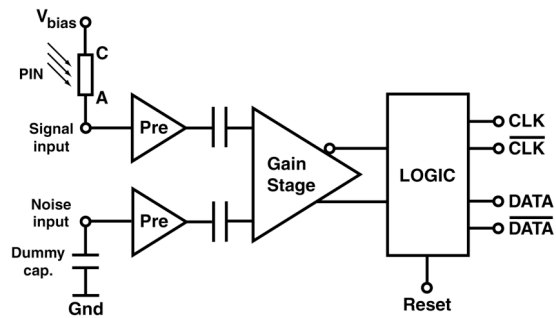

Figure 2. Block diagram of the DORIC circuit.

## 4. Irradiation Studies

We implemented the VDC and DORIC in the standard deep submicron (0.25 µm) CMOS technology. Employing enclosed layout transistors and guard rings [2], this technology was expected to be very radiation hard. We have extensively tested the chips to verify that they satisfy the design specifications [3]. The performance of the chips on opto-boards has been studied in detail. The typical PIN current thresholds for no bit errors are low, < 20 µA.

In the last four years, we have performed four irradiations of the optical electronics using 24 GeV protons at CERN. In June 2004, we irradiated four opto-boards containing complete optical links up to the total fluence of $1.2 \times 10^{15}$ p/cm$^2$ (32 Mrad). The boards were mounted on a shuttle so that they could be pulled back from the beam line each day for annealing of the VCSEL arrays after each dosage of ~ 5 Mrad (5 hours). We observed that the PIN current thresholds for no bit errors were all below 40 μA and remained constant through out the irradiation. The optical power was also monitored and some VCSEL channels lost some of their power after the irradiation due to insufficient time for adequate annealing as shown in Fig. 3. We received the irradiated opto-boards in September 2004 and after extended annealing, the optical power of all VCSEL channels were well above 350 μW, the specification for absolute minimum power after irradiation. This confirms the radiation hardness of the fully populated opto-boards.

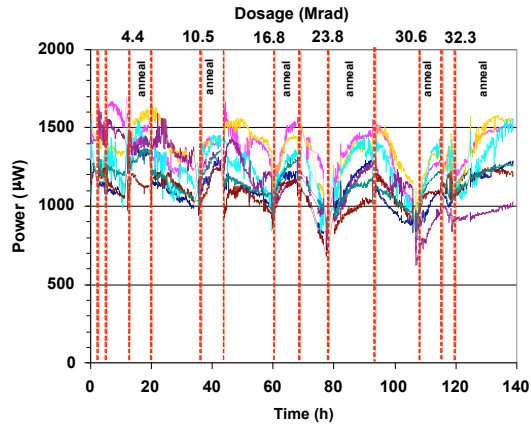

Figure 3. Optical power as a function of time (dosage) in the data channels for one of the opto-boards with seven active links in the shuttle setup.

## 5. Opto-board Production

The opto-board production is governed by a rigorous quality assurance procedure. Each board must passed 72 hours of burn-in at 50°C, followed by 18 hours of 10 thermal cycles between −25 and 50 °C, and then 8 hours of optical and electrical measurements. Our initial plan was not to test the chips before mountingon an opto-board due to the high yield during the pre-production exercise. However, we soon encountered several boards that drawn excessive currents of which we were able to trace the problem quickly to a power to ground short using a high-resolutionthermal imaging camera. Consequently, we tested all chips prior to the mounting on an opto-board. We are also able to recover opto-boards populated with bad chips by stacking new chips on the top. Each reworked opto-board must pass the same rigorous QA procedure but it will be classified as second class for use as a spare. A total of 21 opto-boards are recovered. The production proceeds smoothly as shown in Fig. 4 with high yield, ~ 93%. The production is now completed.

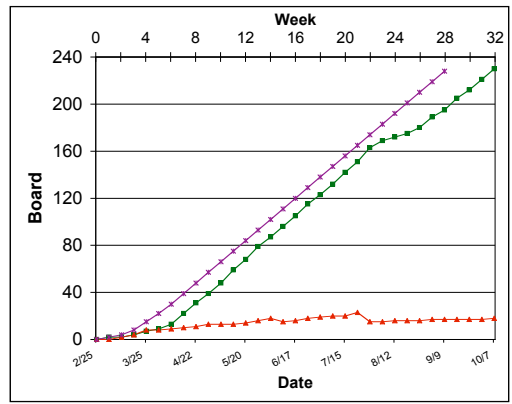

Figure 4: Total number of opto-boards produced as a function of time. The kink at ~ 21$^{st}$ week corresponds to the switch from producing one flavor of opto-broads to the other. Also shown is the expected yield (cross) and total number of failed boards (triangle) which may decrease with time sometimes due to successful rework.

## 6. Summary

We have developed opto-electronics that meet all the requirements for operation in the ATLAS pixel optical link and appear to be sufficiently radiation hard for ten years of operation at the LHC. The production of the opto-boards has proceeded smoothly and is now completed.

### Acknowledgments

This work was supported in part by the U.S. Department of Energy under contract No. DE-FG-02-91ER-40690 and by the German Federal Minister for Research and Technology (BMBF) under contract 056Si74.